\documentclass[aps,pra,twocolumn,showpacs,preprintnumbers,amsmath,superscriptaddress,amssymb,floats,nofootinbib]{revtex4}

\usepackage{graphicx}% Include figure files
\usepackage{dcolumn}% Align table columns on decimal point

\begin{document}

\title{\large \bf Pressure Dependence of Wall Relaxation in Polarized $^3$He Gaseous Cells}
\author{W. Zheng, H. Gao, Q. Ye, Y. Zhang \\
\textit{Triangle Universities Nuclear Laboratory and Department of Physics, Duke University, Durham, NC 27708, USA}}
\noaffiliation

\begin{abstract}
We have observed a linear pressure dependence of longitudinal relaxation time ($T_1$) at 4.2 K and 295 K in gaseous $^3$He cells made of either bare pyrex glass or Cs/Rb-coated pyrex due to paramagnetic sites in the cell wall. The paramagnetic wall relaxation is previously thought to be independent of $^3$He pressure. We develop a model to interpret the observed wall relaxation by taking into account the diffusion process, and our model gives a good description of the data. 
\end{abstract}


\maketitle

Spin polarized $^3$He gas has been widely used in polarized nuclear targets for lepton scattering experiments \cite{Xu} and as a signal source in Magnetic Resonance Imaging (MRI) of lung air space \cite{Middleton2}. Recently, it has also been used in searches of exotic spin-dependent interactions \cite{Petukhov}. These experiments take advantage of large nonequilibrium polarizations of $^3$He obtained through spin-exchange with optically polarized Rb or Rb/K vapor mixture. The production and storage of polarized $^3$He gas crucially depend upon longitudinal relaxation times ($T_1$). Among many factors contributing to the $T_1$ relaxation, the most important ones are the $^3$He dipole-dipole interaction \cite{Newbury}, magnetic field gradient induced relaxation \cite{Cates} and wall relaxation. The least understood and hardest to control among these three effects is the relaxation due to the wall.
 
Although a thorough understanding of the nature of the wall relaxation is lacking, it is widely believed that relaxation rates due to paramagnetic sites in the wall do not depend upon the density of the gas. This is understood in the following way: the wall collision rate per unit area is known as $n\bar{v}/4$, where $n$ is the number density of the gas and $\bar{v}$ is the mean velocity; assuming $\alpha$ is the depolarization probability per collision due to paramagnetic impurities, the relaxation rate $1/T_1$ can be expressed as \cite{Garwin}
\begin{equation}
	\frac{1}{T_1}=\frac{\frac{1}{4}\int\alpha n \bar{v}dS}{\int n dV}=\frac{\alpha\bar{v}S}{4V},
	\label{eq:oldT1}
\end{equation}
where $S$ is the total surface area of the cell, $V$ is the volume of the cell, and $n$ is uniform across the cell. As long as $\alpha$ has no dependence on the gas density or pressure, $T_1$ is also independent of gas pressure.

In this paper, we present our recent $T_1$ measurements on polarized $^3$He cells which show a linear pressure dependence of $T_1$, different from what has been discussed above. In our experiment, the measured $T_1$ is significantly reduced from tens of hours to tens of minutes by just decreasing the pressure of $^3$He gas one hundred times. After excluding dipole-dipole and gradient induced $T_1$ relaxation, the observed pressure dependence can only be explained by the wall relaxation, which is, however, completely opposite to the pressure dependence observed in the ferromagnetic wall relaxation under the weak collision limit \cite{Jacob}. As the cells tested have never been exposed to high fields, ferromagnetic relaxation cannot be the dominant relaxation mechanism, and paramagnetic relaxation is the last candidate to account for the observed $T_1$ relaxation. These $T_1$ measurements have been carried out on cells with surfaces of Cs/Rb-coated pyrex and bare pyrex, and at temperatures 4.2 K and 295K, suggesting that the observed linear pressure dependence is likely a general property of the paramagnetic wall relaxation regardless of the surface and temperature. We also present a model to explain the observed linear pressure dependence by taking into account the diffusion of the spins. This new model also resolves the discrepancies between theories and experiments found in \cite{Chapman,Lusher} and explains a recent finding that polarization and $T_1$ are enhanced by adding $^4$He into the cell \cite{Chen}.

The $T_1$ measurements at 295 K and 4.2 K were carried out using the Free Induction Decay (FID) technique. $^3$He was filled and refilled in 48 mm OD spherical detachable cells to 1 or 2 atm, using a $^3$He/N$_2$ gas handling system. The detachable cell was made of Rb-coated pyrex and had an O-ring valve connected to it through a capillary pyrex tubing with an i.d. of 1.5 mm and a length of 18 cm to restrict gas exchange between the valve and the cell, so that the depolarization from the valve was minimized. The detachable cells were always polarized at either 1 or 2 atm, and then diluted to different pressures (0.025 to 0.43 atm) using different dilution volumes. The pressure in the detachable cell was monitored by a pressure gauge connected between the volume and the cell. 

In the 4.2 K experiment, we measured $T_1$ of $^3$He in cylindrical pyrex cells immersed in liquid $^4$He stored in a dewar. The cylindrical cell had an i.d. of 8 mm and a length of 25 mm. The top of the cell was attached to a thin pyrex tube with an i.d. of 3 mm and a length of 68 cm. The other side of the tube was connected to an O-ring valve outside the dewar, where the detachable cell is mounted. A gas flow restriction (0.8 mm i.d. and 3 mm long) was added to the connection point between the tube and the cell to minimize the gas exchange between them. A dilution volume was also connected for diluted $T_1$ measurements, similar to the 295 K measurement. After dilution, the remaining polarized $^3$He gas in the detachable cell was allowed to diffuse into the cylindrical cell. The pressure in the cylindrical cell can be varied from $6.4\times10^{-4}$ atm to $0.19$ atm. Four cylindrical cells with identical dimensions but different surfaces (two bare pyrex and two Cs-coated pyrex) were used. 

\begin{figure}
	\centering
		\includegraphics[width=0.42\textwidth]{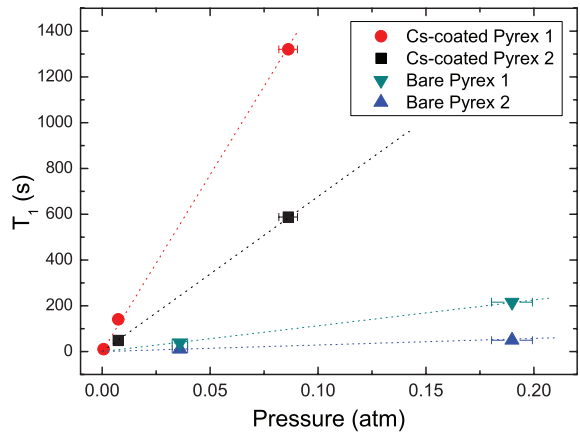}
	\caption{(Color online) $T_1$ of $^3$He in four cylindrical cells at 4.2 K. Two cells are made of bare pyrex (up-triangle and down-triangle) and the other two are made of Cs-coated pyrex (square and circle).}
	\label{fig:4KT1}
\end{figure}

The 295 K $T_1$ measurements were carried out at 39.5 kHz using a homemade FID polarimetry and also at 24 kHz using a commercial polarimetry made by Amersham Health; whereas the 4.2 K measurements were performed at 12 kHz, using Superconducting QUantum Interference Device (SQUID) manufactured by StarCryo. While the 39.5 kHz FID polarimetry has a higher signal-to-noise ratio than the 24 kHz polarimetry, the RF noises from its pre-amplifier are bigger. In all FID measurements, a small tipping angle, resulting in about $1\%$ polarization loss, was applied. This loss was subtracted when extracting $T_1$ from the data. Two different Helmholtz coil pairs were used and the field gradients were measured to be $<2.3$ mG/cm for the 295 K measurement and $<2.2$ mG/cm for the 4.2 K measurement. The gradient induced $T_1$ was more than one thousand hours \cite{Cates}, and was negligible compared to the measured $T_1$. The dipole-dipole induced $T_1$ was calculated~\cite{Mullin,Newbury}, and subtracted from the measured $T_1$ values. Hence, all $T_1$s shown below are due to wall relaxation only.
 
At 4.2 K, when the pressure of the gas is reduced, $T_1$ decreases proportionally (Fig. \ref{fig:4KT1}). From a linear fit of the data (dashed lines), it clearly shows that all the fitted lines pass through the origin, which suggests $T_1\propto p$. For bare pyrex cells, the minimum $T_1$ we measured is 10.1$\pm$0.3 s at $3.6\times10^{-2}$ atm. If the pressure is further reduced, $T_1$ becomes so short that a complete $T_1$ measurement becomes difficult. Cs coating helps increase $T_1$ by more than two orders of magnitude. This allows the pressure to be further reduced to $6.4\times10^{-4}$ atm. However, even with Cs coating, $T_1$ at this pressure is only 10.4$\pm$0.6 s (the first solid circle in Fig. \ref{fig:4KT1}). Therefore, at low pressure, the pressure dependent $T_1$ relaxation at 4.2 K is the dominant relaxation mechanism.  

\begin{figure}
	\centering
		\includegraphics[width=0.42\textwidth]{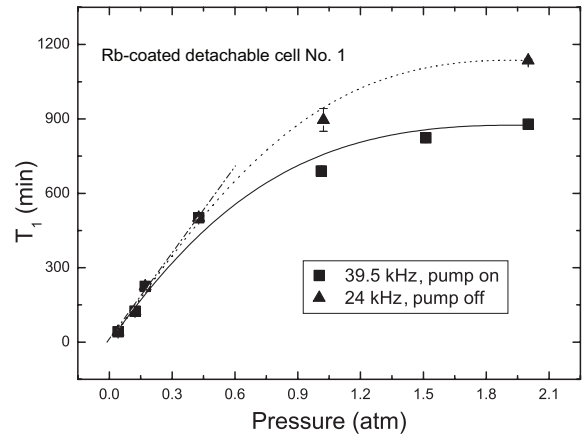}
	\caption{$T_1$ of $^3$He in the Rb-coated detachable cell at 295 K. The dashed line is the linear fit to the first four data points below 0.43 atm. The dotted line is the fit using Eq. (\ref{eq:T1fit}) to the four square points below 0.43 atm and the two triangles points. The solid line is the fit using Eq. (\ref{eq:T1fit}) to all the squares.}
	\label{fig:295KT1-1}
\end{figure}

At 295 K, the Rb-coated detachable cells have $T_1$ on the order of tens of hours. This makes the low pressure $T_1$ measurement easier and also enables us to access the $T_1$-pressure relationship for a different surface and temperature in addition to the 4.2 K data. The detachable cells have never been exposed to fields higher than 30 G. Measurements of $T_1$ have been done to cells before and after degaussing using a commercial demagnetizer, and no changes have been seen within experimental uncertainties. The first detachable cell has $T_1=690\pm21$ min at 1 atm using the 39.5 kHz polarimetry. When the $^3$He pressure is reduced to below 0.43 atm, the measured $T_1$ exhibits a linear pressure dependence (Fig. \ref{fig:295KT1-1}). By fitting the first four points linearly (dashed line), ranging from 0.042 to 0.43 atm, it yields $T_1=1188\times p$ min. This linear dependence does not hold when the pressure is above 0.43 atm, and $T_1$ at 2 atm does not change too much from $T_1$ at 1 atm. This clearly indicates that some other relaxation mechanisms, which are negligible at low pressure, become important at high pressure since the paramagnetic relaxation becomes less pronounced with increasing pressures. In a different experiment, we observed that the continuous RF noise broadcasted by the RF amplifier rendered polarization loss during the $T_1$ measurement \cite{Qiang}. In this experiment, a mechanical pump and a turbo pump were used to maintain the vacuum of the dilution volume throughout the experiment. We indeed observed that the background noise level in the pickup coil increased with the pumps running. Therefore, we repeated the non-diluted 1 atm and 2 atm $T_1$ measurements with all pumps off. The less noisy 24 kHz polarimetry was used in these measurements. The repeated measurements showed an increase of $T_1$ by roughly 200 min and 300 min at 1 atm and 2 atm, respectively. However, in the repeated measurements, $T_1$ still flattens out at 2 atm. As shown in \cite{Jacob2}, all their non-magnetized/demagnetized Rb-coated cells exhibit weak ferromagnetic relaxation behavior. This leads us to believe that our cells are also subject to ferromagnetic relaxation to some extent. As ferromagnetic relaxation has an inverse pressure dependence $T_1\propto 1/p$ \cite{Jacob}, it is negligible at low pressures; whereas, at high pressures, it becomes more prominent and therefore comparable to the paramagnetic relaxation. This could explain the flattening behavior of $T_1$. Hence, the measured $T_1$ relaxation can be attributed to three mechanisms, expressed as
\begin{equation}
	\frac{1}{T_1}=\frac{1}{c_1p}+\frac{1}{c_2}+\frac{p}{c_3}.
	\label{eq:T1fit}
\end{equation}
The first term is the paramagnetic wall relaxation, which depends on pressure linearly ($T_1=c_1p$); the second term is RF noise-induced relaxation, which has no pressure dependence ($T_1=c_2$); the last term is the ferromagnetic relaxation, which inversely depends on pressure ($T_1=c_3/p$) \cite{Jacob,Schmiedeskamp}. Using Eq. (\ref{eq:T1fit}) to fit the 39.5 kHz data in Fig. \ref{fig:295KT1-1} (solid line), one obtains $c_1=1288\pm110$ min/atm, $c_2=2754\pm251$ min and $c_3=5082\pm1641$ min$\cdot$atm. If the two repeated measurements together with the four data points below 0.43 atm were used for the fit (dotted line), $c_2$ changes to $c_2\geq10086$ min, indicating $1/c_2$ is zero within fitting errors. This suggests that the RF noise in the repeated measurements is negligible.

\begin{figure}
	\centering
		\includegraphics[width=0.42\textwidth]{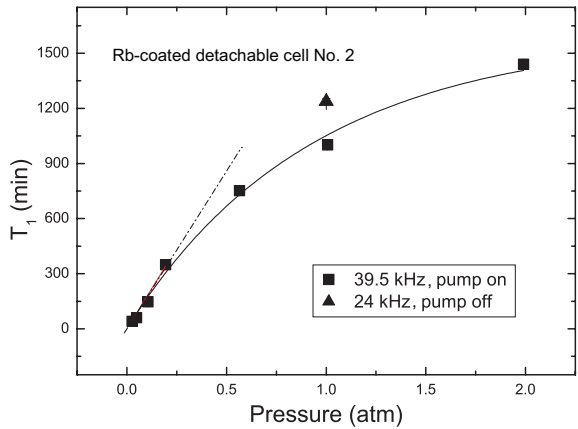}
	\caption{$T_1$ of $^3$He in the second detachable cell which has longer $T_1$ at 1 atm. The dashed line is the linear fit to the four data points with pressure below 0.25 atm. The solid line is a fit using Eq. (\ref{eq:T1fit}) to all 39.5 kHz data labeled as squares.}
	\label{fig:295KT1-2}
\end{figure}

We also tested another Rb-coated detachable cell, which has a longer $T_1$ at 1 atm ($T_1=1001\pm11$ min), measured by the 39.5 kHz polarimetry. This cell also shows the linear pressure dependence with pressure below 0.25 atm (Fig. \ref{fig:295KT1-2}); and beyond this pressure, $T_1$ starts to flatten out as well. Fitting the data to Eq. (\ref{eq:T1fit}) yields $c_1=1757\pm71$ min/atm, $c_2=2959\pm362$ min and $c_3\geq10113$ min$\cdot$atm. $c_2$ of this cell is comparable to $c_2$ of the first one as expected, since the RF noise-induced $T_1$ should be independent of which cell is used. Both $c_1$ and $c_3$ are larger than those of the first cell, which suggests that both paramagnetic and ferromagnetic wall relaxations of the second cell are weaker than those of the first one. The 1 atm $T_1$ was also measured using 24 kHz polarimetry with all pumps off. $T_1$ in this case increases by roughly 236 min. 

Once ferromagnetic wall relaxation and RF noise-induced relaxation are excluded, the $T_1$ relaxation measurements at both 295 K and 4.2 K clearly show the linear pressure dependence, which is contradictory to Eq. (\ref{eq:oldT1}). The derivation of Eq. (\ref{eq:oldT1}) implicitly assumes ballistic collisions between spins and the wall, which is only true when spins are in the vicinity of the wall. When far away from it, spins move in a diffusive manner. Hence, the effective speed, at which it moves to the wall, is much slower than its thermal velocity. As majority of the spins are not close to the wall, a more appropriate model to describe the wall relaxation should take into account the diffusion process. In \cite{Chupp}, Chupp \textit{et al.} used the diffusion equation to calculate the spatial distribution of $^3$He polarization inside a high pressure double cell system. We will also use the diffusion equation, together with depolarizing boundaries, to describe the surface relaxation of $^3$He. It should be noted that ferromagnetic relaxation does not fit into this surface relaxation model because it happens not only on the surface but also in the vicinity of the surface. Since ferromagnetic impurities produce much stronger dipole field than paramagnetic impurities, spin can be depolarized even it does not have a contact with the surface. In other words, ferromagnetic relaxation actually occurs in a region adjacent to the surface impurities. In the strong collision limit of the ferromagnetic relaxation, the dipole field is so strong that the adjacent relaxation region extends to the entire cell. In this case, the relaxation rate converges to the gradient-induced relaxation \cite{Jacob,Schmiedeskamp}. 

Let $\rho(r,t)$ represent the polarization of $^3$He gas inside a spherical cell as a function of position $r$ measured from the center of the cell and time $t$. The diffusion equation of the polarization $\rho$ is written as
\begin{equation}
	D\nabla^2\rho=\frac{\partial\rho}{\partial t},
	\label{eq:diffeqn}
\end{equation}
where $D$ is the diffusion constant of $^3$He gas. Since spins lose their polarization only at the surface with probability $\alpha$, the boundary condition is written as
\begin{align}
	&\frac{\partial\rho(r,t)}{\partial r}|_{r=0}\neq\infty
	\label{eq:bc1} \\
	&\frac{\partial\rho(r,t)}{\partial r}|_{r=R}=-\alpha\rho(R,t),
	\label{eq:bc2}
\end{align}
where $R$ is the radius of the cell. The solution is
\begin{equation}
	\rho(r,t)=\sum^{\infty}_{k=1}A_kj_0(\frac{x_k r}{R})\exp(\frac{-x_k^2Dt}{R^2}),
\end{equation}
where $j_0$ is the zeroth order spherical Bessel function, $x_k$ is the k$^{th}$ root of Eq. (\ref{eq:bc2}), which can be re-written as
\begin{equation}
	x_kj^{\prime}_{0}(x_k)+\alpha Rj_0(x_k)=0,
	\label{eq:xn}
\end{equation}
and $A_k$ is determined by
\begin{equation}
	A_k=\frac{\int^{R}_{0}\rho(r,0)j_0(\frac{x_k r}{R})r^2dr}{\int^{R}_{0}j^2_0(\frac{x_k r}{R})r^2dr}.
\end{equation}
Since terms other than $k=1$ vanish quickly, only $k=1$ term contribute to the polarization and $T_1$ is written as
\begin{equation}
	1/T_1=\frac{x_1^2D}{R^2}\propto\frac{x_1^2}{nR^2}\propto\frac{x_1^2}{pR^2}.
	\label{eq:newT1}
\end{equation}
The second and third proportionalities use the fact that $D\propto 1/n\propto 1/p$. Therefore, the paramagnetic wall relaxation rate $1/T_1$ does not only depend upon the depolarization probability $\alpha$ (implicitly through $x_1$) but also on the diffusion constant $D$. Chen \textit{et al.}\cite{Chen} found that $T_1$ of their $^3$He cells are increased by filling $^4$He into the cell. This can be explained by the decrease of $^3$He diffusion constant due to the presence of $^4$He in the cell. It should be noted that Eq. (\ref{eq:diffeqn}) can also be used in situations other than pure $^3$He gas, for instance $^3$He in superfluid $^4$He. In this case, $D$ is the diffusion constant of $^3$He in superfluid $^4$He \cite{Murphy,Lamoreaux}, so Eq. (\ref{eq:diffeqn}) and the following arguments are still valid. Another observation on Eq. (\ref{eq:newT1}) is that $1/T_1$ has a quadratic dependence on the surface-to-volume ratio, $1/T_1\propto 1/R^2\propto (S/V)^2$ (spherical cell). This quadratic dependence, instead of the linear dependence in Eq. (\ref{eq:oldT1}), suggests that surface to volume ratio has bigger influence on paramagnetic relaxation $T_1$ than it is previously believed.

Low pressure cells (a few torr) used in Metastability Exchange Optical Pumping (MEOP) have been reported to have long $T_1$ from several hours to tens of hours \cite{Colegrove,Heil,Cheron,WChen}. These include both valved cells and permanently sealed cells; and they are usually made of aluminosilicate glass or Cs-coated pyrex with much better glass cleaning process than what we did to our detachable cells. No apparent pressure dependence was observed in the valved MEOP cells, except that, at low enough pressures, the gradient-induced relaxation dominates. This seems to be inconsistent with the pressure dependence we observed. However, these cells usually have hundreds or even thousands of hours of $T_1$ at 1 atm, considerably longer than our cells (less than 20 hours). Rather than the paramagnetic relaxation, $T_1$ of those MEOP cells are likely dominated by other relaxation mechanisms, such as the ferromagnetic relaxation and the dipole-dipole relaxation.

$T_1$ of ferromagnetic wall relaxation has an inverse linear pressure dependence in the weak collision limit, defined as $\omega_0\tau<<1$, where $\omega_0$ is the spin precession frequency and $\tau$ is the interaction time of spin with a magnetic site on the surface \cite{Jacob,Schmiedeskamp}. In the strong collision limit ($\omega_0\tau>>1$), $T_1$ becomes linearly dependent on pressure, which in fact can be understood by the gradient-induced relaxation \cite{Jacob}. As the experimental conditions of both 4.2 K and 295 K measurements are clearly not in the strong collision limit which requires large ferromagnetic site on the surface, and our cells have never been exposed to high fields, the observed pressure dependence cannot be explained by the ferromagnetic relaxation in the strong collision limit.

Chapman and Richards \cite{Chapman} also observed the linear pressure dependence of $T_1$ in $^3$He at 4.2 K. They use Eq. (\ref{eq:oldT1}) to describe their findings and the pressure dependence was ascribed to the pressure dependence of $\alpha$ using a two-phase model with $^3$He in the absorbed phase with a shorter $T_1$ when a complete monolayer was formed on the surface, and $^3$He in the bulk phase which has much longer $T_1$. However, this two-phase picture cannot explain the pressure dependence seen in the experiment by Lusher \textit{et. al.}~\cite{Lusher}, in which only a partial monolayer was formed. The binding energies $W$ between $^3$He spin and a specific surface determines when a complete monolayer will be formed. For bare pyrex glass, the binding energy is around 100 K \cite{Lusher}. For Cs and Rb-coated cells, the binding energy are 2.3 K and 2.8 K \cite{Deninger2,Cheng}, respectively. In our 4.2 K measurement, for sure a complete monolayer of $^3$He is formed on the bare pyrex; whereas, in the Cs-coated cell, it is certain that only a partial monolayer is formed. As the pressure dependence is observed in both cases, it further demonstrates that Eq. (\ref{eq:oldT1}) and the two-phase picture are inadequate to describe the experimental data, and the diffusion process is essential to the wall relaxation.   

In summary, we conclude that the linear pressure dependence observed in our $T_1$ measurements is associated with the paramagnetic wall relaxation. This pressure dependence originates from the diffusion process of $^3$He spins and can be well described by the diffusion equation. It also suggests that it is vital to control the paramagnetic wall relaxation when the diffusions are fast.

We thank J. Rishel for making all the cells reported in this work, and Tom Gentile, Jian-guo Liu and Mike Snow for helpful discussions. This work is supported by the U.S. Department of Energy under contract number DE-FG02-03ER41231 and Duke University.

\end{document}